\documentclass[10pt]{article}
\usepackage{amsmath}
\usepackage{graphicx,psfrag,epsf}
\usepackage{enumerate}
\usepackage[square,numbers]{natbib}
\usepackage{url} % not crucial - just used below for the URL 
\usepackage[skip=0.8\baselineskip,width=1.25\textwidth,font=small]{caption}

\usepackage{graphicx}
\usepackage{tabularx}
\usepackage[utf8]{inputenc}
\newcommand{\blind}{1}
\newcommand{\htwo}{\ensuremath{\text{H}_2} }
\newcommand{\heh}{\ensuremath{\text{HeH}^+} }
\newcommand{\lih}{\ensuremath{\text{LiH}} }
\newcommand{\ethylene}{\ensuremath{\text{C}_2\text{H}_4} }
\usepackage{authblk}
\begin{document}

\def\spacingset#1{\renewcommand{\baselinestretch}%
{#1}\small\normalsize} \spacingset{1}

%%%%%%%%%%%%%%%%%%%%%%%%%%%%%%%%%%%%%%%%%%%%%%%%%%%%%%%%%%%%%%%%%%%%%%%%%%%%%%

\if1\blind
{
  \title{\bf Statistical learning method for predicting density-matrix based electron dynamics }
  \author[1,2]{Prachi Gupta\thanks{pgupta11@ucmerced.edu}}
  \affil{Chemistry and Biochemistry, University of California, Merced}
  \author[2]{Harish~S. Bhat\thanks{hbhat@ucmerced.edu}} 
  \affil{Applied Mathematics, University of California, Merced}
  \author[1]{Karnamohit Ranka}
  %\affil{Chemistry and Biochemistry, University of California Merced}
  \author[1]{Christine~M. Isborn\thanks{cisborn@ucmerced.edu}} 
  %\affil{Applied Mathematics, University of California Merced}
    
  \maketitle
} \fi

\if0\blind
{
  \bigskip
  \bigskip
  \bigskip
  \begin{center}
    {\LARGE\bf Title}
\end{center}
  \medskip
} \fi

\bigskip

\abstract{We develop a statistical method to learn a molecular Hamiltonian matrix from a time-series of electron density matrices. 
We extend our previous method to larger molecular systems by incorporating physical properties to reduce dimensionality, while also exploiting regularization techniques like ridge regression for addressing multicollinearity. With the learned Hamiltonian we can solve the Time-Dependent Hartree-Fock (TDHF) equation to propagate the electron density in time, and predict its dynamics for field-free and field-on scenarios. We observe close quantitative agreement between the predicted dynamics and ground truth for both field-off trajectories similar to the training data, and field-on trajectories outside of the training data. 
%which is evidenced by the least square estimate providing a good estimate to training set but not one the test set.  
}
\\

\noindent%
\vfill

% \jnlcitation{\cname{%
% \author{Williams K.}, 
% \author{B. Hoskins}, 
% \author{R. Lee}, 
% \author{G. Masato}, and 
% \author{T. Woollings}} (\cyear{2016}), 
% \ctitle{A regime analysis of Atlantic winter jet variability applied to evaluate HadGEM3-GC2}, \cjournal{Q.J.R. Meteorol. Soc.}, \cvol{2017;00:1--6}.}

% \footnotetext{\textbf{Abbreviations:} ANA, anti-nuclear antibodies; APC, antigen-presenting cells; IRF, interferon regulatory factor}

\section{Introduction}\label{sec1}

Predicting the dynamic electronic properties of a molecular system is essential to understanding phenomena such as charge transfer and response to an applied laser field. The time-dependent Schr\"{o}dinger equation (TDSE) governs the time evolution of a quantum electronic system. %The electronic time-dependent Schrodinger equation (TDSE) governs the time evolution of a quantum electronic system,  
% \begin{equation}
% \label{eqn:TDSE}
% i \frac{\partial \Psi(\mathbf{r_1,r_2...,r_N},t)}{\partial t} = \hat H (\mathbf{r_1,r_2...,r_N},t) \Psi(\mathbf{r_1,r_2...,r_N},t)
% \end{equation}
% where $\Psi$ is the many-body wave function , $\mathbf{r_N}$ is the coordinates of electron $N$, and $\hat H$ is the electronic Hamiltonian operator that is the sum of kinetic and potential energy operators of the electronic system. For a molecule, the electronic Hamiltonian operator is: 
% \begin{equation}
% \label{eqn:Hamoperator}
% \hat{H}(\mathbf{r},t) = \sum_{i}^{N_e}\left[ -\frac{\nabla^2}{2} - \sum_A \frac{Z_A}{| \mathbf{r}_i - \mathbf{R}_A|}  \right]  + \sum_{i}^{N_e} \sum_j^{N_e}\frac{1}{|\mathbf{r}_i-\mathbf{r}_j|} + V_{\mathrm{ext}}(\mathbf{r},t)
% \end{equation}
% where $N_e$ is the number of electrons,  and $Z_A, R_A$ denote the nuclear charge and position of nucleus A. In (\ref{eqn:Hamoperator}), the first term is the one-electron term consisting of the electronic kinetic energy and electron-nuclear attraction operators. The second term represents the two-electron Coulombic repulsion between electrons. The third term $V_{\text{ext}}$ is an external potential, which in our study will consist of an applied electric field.  Within the dipole approximation, this one-electron potential can be written as $V_{\text{ext}} = \sum_{i}^{N_e} \mathbf{E}(\mathbf{r}, t) \cdot \mu(\mathbf{r_i})$, where  $\mathbf{E}(\mathbf{r}, t)$ is the external field in the $r^{th}$ Cartesian direction and $\mu(\mathbf{r})$ is the dipole moment operator.  
Using the time-dependent density operator within a finite-dimensional basis yields the Liouville-von Neumann equation:% within the density matrix formulation as,
\begin{equation}
\label{eqn:LiNe}
i\frac{d \mathbf{P}(t)}{d t} =  \big [   \mathbf{H}(t), \mathbf{P}(t) \big].
\end{equation}
Here $\mathbf{P}(t)$ and $\mathbf{H}(t)$ denote the time-dependent electron density and Hamiltonian matrices in orthonormal bases, respectively, and the square brackets denote a commutator: for any square matrices $\mathbf{A}$ and $\mathbf{B}$, the commutator is $[\mathbf{A},\mathbf{B}] = \mathbf{A}\mathbf{B} - \mathbf{B} \mathbf{A}$.

The many-body problem given by  \eqref{eqn:LiNe} can only be solved for simple systems, such as those with very few electrons within a small basis. 
% Therefore, chemists often use Time-Dependent Density Functional Theory (TD-DFT), which reformulates the TDSE in terms of the electron density instead of the wave function \cite{Runge1984Density-functionalSystems}. Key to this theory is the introduction of a density dependent exchange correlation potential ($v_{XC}(\mathbf{r},t)$) that accounts for quantum electron-electron many-body Coulombic interactions not captured from the classical (mean-field) Coulomb contribution.  Within TDDFT, the accuracy is limited by the accuracy of approximations to $v_{XC}(\mathbf{r},t)$ as its exact form is unknown. Analyzing errors in TDDFT simulations due to approximations for $v_{XC}(\mathbf{r},t)$ and developing new approximations is an active area of research.\cite{Habenicht2014Two-electronTheory, Provorse2015, Maitra16_220901} Machine learning methods have increasingly been used in quantum chemistry for developing predictive models for potential energy surfaces \cite{Artrith2011,Behler2007,Smith2017,braams2009permutationally}, finding functionals for DFT \cite{Snyder2012,li2016,Suzuki2020}, and learning Hamiltonians \cite{Fujita2018}. 
%Machine learning methods have increasingly been used in quantum chemistry for developing predictive models for potential energy surfaces \cite{Artrith2011,Behler2007,Smith2017,braams2009permutationally}, finding functionals for DFT \cite{Snyder2012,li2016,Suzuki2020} and learning Hamiltonians \cite{Fujita2018}.
Hartree-Fock (HF) theory is a simplified mean field approach in which the many-body wave function is approximated using an anti-symmetrized product of single particle orbitals. Applying this approximation to the Hamiltonian for   \eqref{eqn:LiNe} produces two-electron terms that are given by Coulomb and exchange operators, and thus a Hamiltonian that is now density dependent $\mathbf{H}(\mathbf{P})$.  Using this HF Hamiltonian, sometimes called the Fock matrix within HF theory, with \eqref{eqn:LiNe} yields the time-dependent HF (TDHF) equation, which, along with time-dependent density functional theory (TDDFT), is often used for simulating electron dynamics, 
\begin{equation}
\label{eqn:eq1}
i\frac{d \mathbf{P}(t)}{d t} =  \big [   \mathbf{H}(\mathbf{P},t), \mathbf{P}(t) \big].
\end{equation}

Here, using TDHF training data, we address the problem of \emph{learning the field-free Hamiltonian matrix $\mathbf{H}(\mathbf{P})$ from time series observations of electron densities $\mathbf{P}(t)$}. For the field-free trajectory, {\it i.e.}, when the Hamiltonian contains no explicit time-dependence, $\mathbf{H}$ is a complex Hermitian matrix function of $\mathbf{P}$, which is also complex and Hermitian. Therefore, $\mathbf{H}$ and $\mathbf{P}$ are completely determined by their upper triangular elements. Both matrices can be represented by vectors that contain the real and imaginary components of their upper triangular parts. Using these vector representations for $\mathbf{H}$ and $\mathbf{P}$, we develop a statistical model for $\mathbf{H}$. This model is linear in its parameters $\beta$; in the vector representation, the model Hamiltonian is also a linear function of electron density matrix elements.

To fit the model, we minimize a loss function that measures the squared Frobenius norm between the left- and right-hand sides of \eqref{eqn:eq1}, evaluated on training data. This data consists of time series of electron density matrices $\mathbf{P}(t)$ and their time-derivatives $d \mathbf{P} / dt$ computed via centered differencing.  The loss function depends on its parameters through the Hamiltonian. Since we use a linear model for the Hamiltonian, our loss function is quadratic in the model parameters $\beta$. Therefore, to minimize the loss and fit the model, we must solve a least squares problem.  Equipped with the Hessian $H$ and gradient $g$ of the loss function, the solution to this problem reduces to that of $H\beta = g$.  For small systems, we can carry this out effectively, using automatic differentiation to compute $H$ and $g$.

However, this approach does not scale well to larger molecular systems and results in prohibitively large training times, the majority of which is required for computation of the Hessian matrix. To address this, we develop a data science framework that scales to larger molecules and larger basis sets than in our previous work \cite{bhat2020machine}. Here, we use dimensionality-reduction techniques based on degrees of freedom in the density matrix and properties of the HF Hamiltonian. Another challenge for large systems with symmetry is that the Hamiltonian model does not extrapolate well to the field-on case because the Hessian matrix has $0$ eigenvalues, leading to multicollinearity. To resolve this challenge we use ridge regression. Ridge regression places a constraint on the model parameters by adding a penalty to the loss function. 

To train the Hamiltonian model we use time series of density matrices generated with no external perturbations. Using the learned Hamiltonian, we propagate forward in time to obtain a field-free trajectory. To compute a field-on trajectory, we add a time-dependent external perturbation to the learned Hamiltonian and propagate forward in time.  We find that the learned field-free Hamiltonian can be used to propagate electron dynamics in both field-free and field-on conditions, yielding results that closely match those obtained via ground truth Hamiltonians.

Our overarching goal is to learn a potential/Hamiltonian for TDDFT to simulate more accurate electron dynamics. Key to this theory is the introduction of a density dependent exchange correlation potential $v_{XC}(\mathbf{r},t)$ that accounts for quantum electron-electron many-body Coulombic interactions not captured from the classical (mean field) Coulomb contribution. However, the exact form of the exchange correlation portion of the Hamiltonian is unknown. Therefore, our goal is to first develop a method to learn a known, more approximate density-dependent Hamiltonian, like  that used in time-dependent Hartree-Fock (TDHF) theory \cite{Rubio11_book, Isborn16_739, Maitra16_220901,LiGovind2020}. This work provides the methodological development for a framework that seeks to model the Hamiltonian and use it to predict the dynamics of the system. This work sets us on a pathway towards developing a novel statistical/machine learning method for more complex theories for predicting electron dynamics.

%  In doing so, we deviate from the traditional time series prediction approach of using more popular Neural Network based methods such as Recurrent Neural Networks, which are difficult to train, computationally expensive, and do not extrapolate well beyond the training set.

\section{Methods}\label{sec2}

\subsection{Generating Data}
In this paper, we predict electron dynamics for six molecular systems: \htwo in the 6-31G basis set (two electrons in 4 basis functions), \heh in the 6-31G and 6-311++G$^{**}$ basis sets (two electrons in 4 and 14 basis functions, respectively), \lih in the 6-31G and 6-311++G$^{**}$ basis sets (four electrons in 11 and 29 basis functions, respectively), and \ethylene in the STO-3G basis set (16 electrons in 14 basis functions). Note that each molecular orbital created from a linear combination of these atomic orbital basis functions is doubly occupied. We build off of our previous work that developed models for the simpler systems, \htwo, \heh and \lih in the STO-3G basis set (two electrons and two basis functions) \cite{bhat2020machine}.

For each molecular system, we apply standard electronic structure methods to compute the ground truth field-free Hamiltonian/Fock matrix $\mathbf{H}(\mathbf{P})$ and variationally determine the ground state electron density matrix $\mathbf{P}$. Our initial condition at $t=0$ is either field-free with $\mathbf{P}(0)$ determined from solving for the electron density in the presence of an applied electric field (a delta-kick perturbation at $t=0$), or $\mathbf{P}(0)$ is the ground state electron density and we apply the field during propagation (see below).  We then numerically solve (\ref{eqn:eq1}) to generate an electron dynamics trajectory $\mathbf{P}(t)$, recording the data at temporal resolution $\Delta t = 0.08268$ a.u., propagating with the modified midpoint unitary transformation method \cite{Li2005, SSL2007}. These steps were performed using a modified version of the Gaussian electronic structure code \cite{GaussianDV}. We generate two data sets for each molecular system:
\begin{enumerate}[1.]
\item \textbf{Field-free trajectory}: The initial density matrix in the presence of an electric field is calculated. Using this initial condition as the delta kick perturbation and then without applying any external perturbation during propagation, a trajectory is produced. A part of this trajectory, i.e., density matrices $P(t_j)$ where $t_j = j\Delta t \text{ for } 2\leq j \leq N$, is used for training and another part of this trajectory for $ N+1\leq j \leq M$ is used as a validation set.

\item \textbf{Field-on trajectory}: The initial ground state density matrix without any perturbation is calculated.  An external forcing term $\mathbf{V}_{\text{ext}}(t) =  E_z sin(\omega t)\mu_z$ is applied during propagation, where $E_z$ is the applied electric field in the z direction (along the main molecular bond axis), $\omega$ is the electric field frequency, and $\mu_z$ is the $z$ component of the molecular dipole moment. For this study, the electric field is turned on for one cycle ($3.55 \text{fs}$ = $147 \text{a.u.}$) at $t=0$, with $\omega = 0.0428$ a.u (an off-resonant frequency corresponding to the neodymium-YAG laser) and $E_z = 0.05$ a.u. We test our learned Hamiltonian against this field-on trajectory; field-on trajectories are never used during the training process.  

\end{enumerate}

\subsection{Statistical Learning}
\label{sect:statlearning}
Our aim is to learn the molecular Hamiltonian $\mathbf{H}(\mathbf{P})$, which is a Hermitian matrix-valued function of the Hermitian density matrix $\mathbf{P}$ as in (\ref{eqn:eq1}).  Since $\mathbf{H}$ and $\mathbf{P}$ are Hermitian, they are completely determined by their upper-triangular components. We split $\mathbf{H}$ and $\mathbf{P}$ into real and imaginary matrices and then flatten and combine the upper-triangular parts of each matrix into corresponding real vectors. Let $\mathbf{h}$, $\mathbf{p}$ denote real column vectors that contain the real and imaginary parts of the upper-triangular portions of the complex matrices $\mathbf{H}$, $\mathbf{P}$. Let tildes denote statistical models---to be clear, $\widetilde{\mathbf{H}}$ is the model Hamiltonian, different from the true Hamiltonian $\mathbf{H}$.  As in \cite{bhat2020machine}, we use a linear model and squared loss
\begin{align}
    \label{eqn:hammodel}
    \widetilde{\mathbf{h}}(\mathbf{p}) &= \beta_0 + \beta_1 \mathbf{p} \\
\label{eqn:sseloss}
    \mathcal{L}( \beta ) &= \sum_{j=1}^{N-1} \biggl\| i  \frac{\mathbf{P}_{j+1} - \mathbf{P}_{j-1}}{2 \Delta t}
    -  \big [   \widetilde{\mathbf{H}}_j , \mathbf{P}_j \big] \bigg\|_F^2,
\end{align}
where  $\beta = (\beta_0, \beta_1)$, $\mathbf{P}_j = \mathbf{P}(t_j)$, $\widetilde{\mathbf{H}}_j = \widetilde{\mathbf{H}}(\mathbf{P}(t_j))$, and $t_j = j \Delta t$.  The loss function quantifies the mismatch between the left- and right-hand sides of (\ref{eqn:eq1}), with the time-derivative approximated by a centered-difference quotient.  To train, we compute $\beta$ that minimizes $\mathcal{L}$ on the training data:
\begin{equation}
\label{eqn:lstsqprob}
    \beta^* \in \underset{\beta}{\arg\min} \{\mathcal{L}(\beta)\}.
\end{equation} 
This is a least squares problem. Let $Q$ denote the Hessian of the loss with respect to $\beta$.  Let $c = \nabla_{\beta}\mathcal{L}( 0 )$, the gradient of the loss with respect to $\beta$, evaluated at $\beta=0$.  To solve (\ref{eqn:lstsqprob}), we can take the gradient of the loss function and set it to 0. This results in the normal equations, which we can write in terms of the Hessian and gradient of the loss (see Appendix \ref{app1} for details):
\begin{align}
    \label{eq5}
    Q \beta = -c.
\end{align}
We briefly explain the meaning of the loss $\mathcal{L}$ by asking the hypothetical question: if $\mathbf{P}(t_j)$ refers to ground truth electron density matrices in our training data, what does it mean for the loss function to vanish?  Consider the following equation, which defines a \emph{one-step prediction} of $\mathbf{P}_{j+1}$:
\begin{equation}
\label{eqn:onesteppred}
\widetilde{\mathbf{P}}_{j+1} = \mathbf{P}_{j-1} - 2 i \Delta t -  \big [   \widetilde{\mathbf{H}}(\mathbf{P}_j), \mathbf{P}_j \big]
\end{equation}
For $\mathcal{L}(\beta)$ to vanish, for each $j$, we must be able to insert the true values of $\mathbf{P}_{j-1}$ and $\mathbf{P}_{j}$ into the right-hand side of (\ref{eqn:onesteppred}) and obtain a predicted $\widetilde{\mathbf{P}}_{j+1}$ that perfectly matches the true $\mathbf{P}_{j+1}$.  In short, the loss measures the deviation from perfect one-step or local prediction via (\ref{eqn:onesteppred}), across the entire training time series. We use the loss $\mathcal{L}$ as a proxy for the true metric of interest, which is long-term propagation error \eqref{eqn:properror}.  Direct or adjoint-based minimization of \eqref{eqn:properror} can in principle be used to solve for $\beta$; however, this will be much more computationally expensive than our approach.

As described above, the training data consists of field-free trajectories.
For each molecular system, we train using time series of density matrices $\mathbf{P}(t_j)$ where $t_j = j\Delta t \text{ for } 2\leq j \leq N$ obtained from the field-free trajectory to ensure that this learned Hamiltonian does not depend on an external field. We do not use the first two time steps of the trajectory since these time steps have large values of ${d\mathbf{P}}/{dt}$, a consequence of the delta-kick initial condition.

The solution to  (\ref{eq5}) results in the statistical estimates $\beta$ and the molecular Hamiltonian can then be determined using (\ref{eqn:hammodel}). We tested the model for small molecules in small basis sets (up to 6$\times$ 6 in dimension for the complex Hamiltonian).  When we sought to extend this approach to more complex molecular systems, we encountered two main problems: (i) training times were unacceptably large due to automatic differentiation, and (ii) propagation results were inaccurate. To solve (i), we coded the gradient and Hessian of the loss (\ref{eqn:sseloss}) ourselves, leveraging parallelization---see Appendix \ref{sect:gradHess} for details.  To solve (ii), we applied  \emph{dimensionality reduction} and \emph{ridge regression}, which we now detail in turn.

\subsubsection{Dimensionality Reduction}
We consider diatomic molecules in the 6-31G and 6-311++G** bases and the larger molecule \ethylene in the small STO-3G basis.  Let $N$ denote the dimension of the density and Hamiltonian matrices for each molecule in a given basis set. For larger basis sets or larger molecules, $N^2$ increases dramatically; see Table \ref{tab1}.  Our initial implementation leads to a na\"ive version of (\ref{eqn:hammodel}) in which $\mathbf{p}$ is of size $N^2 \times 1$ and hence $\beta$ is an $(N^2 + 1) \times N^2$ matrix of regression coefficients.  We employ two tactics to reduce the dimensionality of $\beta$.  First, we split (\ref{eqn:hammodel}) into two separate models, such that the parts of $\widetilde{\mathbf{h}}$ that correspond to \emph{real} (respectively, \emph{imaginary}) components of $\widetilde{\mathbf{H}}$ depend only on the \emph{real} (respectively, \emph{imaginary}) components of $\mathbf{P}$.  This splitting, which can be justified based on physical properties of the Hartree-Fock Hamiltonian, was not present in our prior work \cite{bhat2020machine}. At time $t_j = j \Delta t$, the true field-free Hamiltonian in the AO basis is,
\begin{equation}
\label{eqn:ffham}
\mathcal{H}^j = \mathcal{K} - \mathcal{N} + \mathcal{V}(\mathcal{P}^j).
\end{equation}
Here $\mathcal{K}$ is the kinetic energy matrix, $\mathcal{N}$ is the electron-nuclear energy matrix, and $\mathcal{V}$ is the density dependent combination of Coulomb and exchange matrices. Let $\mathcal{V}^j = \mathcal{V}(P^j)$, then for $u \leq v$,
\begin{equation}
\mathcal{V}^j_{u,v} = \sum_{l, s} 2 \mathcal{P}^j_{l,s} \left( \mathcal{E}_{u,v,l,s} - \frac{1}{2} \mathcal{E}_{u,l,v,s} \right),
\end{equation}
where $\mathcal{E}$ is a four-index tensor in the Coulomb and exchange calculations. Because this tensor is real, the real elements of the Hamiltonian depend on the real elements of the density matrix and the imaginary elements of the Hamiltonian depend on the imaginary elements of the density matrix.  The second tactic used to reduce dimensionality is that when forming the flattened vector representation $\mathbf{h}$, we retain only those entries of $\mathbf{H}$ where the corresponding entries of $\mathbf{P}$ are not identically zero \cite{bhat2020machine}. For these linear or flat molecular systems, elements are identically zero due to the molecular symmetry, e.g. if they are constructed from orthogonal basis functions. In this way, for the largest problem under consideration, we reduce $\beta$ from $842 \times 841$ to $226\times225$, reducing the number of coefficients by a factor $>13.9$.

\subsubsection{Ridge Regression}
\begin{figure}[t]
\centering
\includegraphics[width=3.2in]{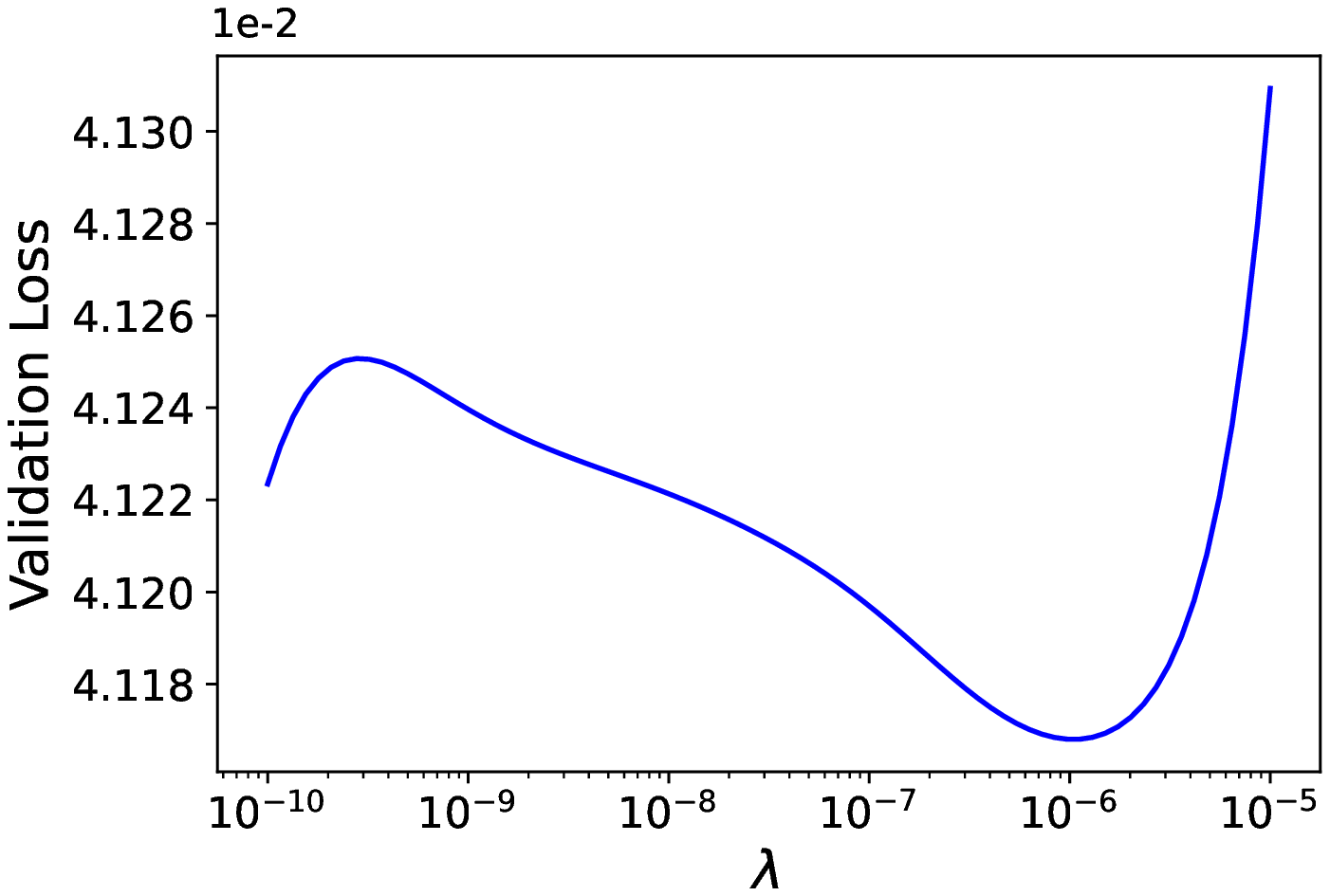}
\includegraphics[width=3.1in]{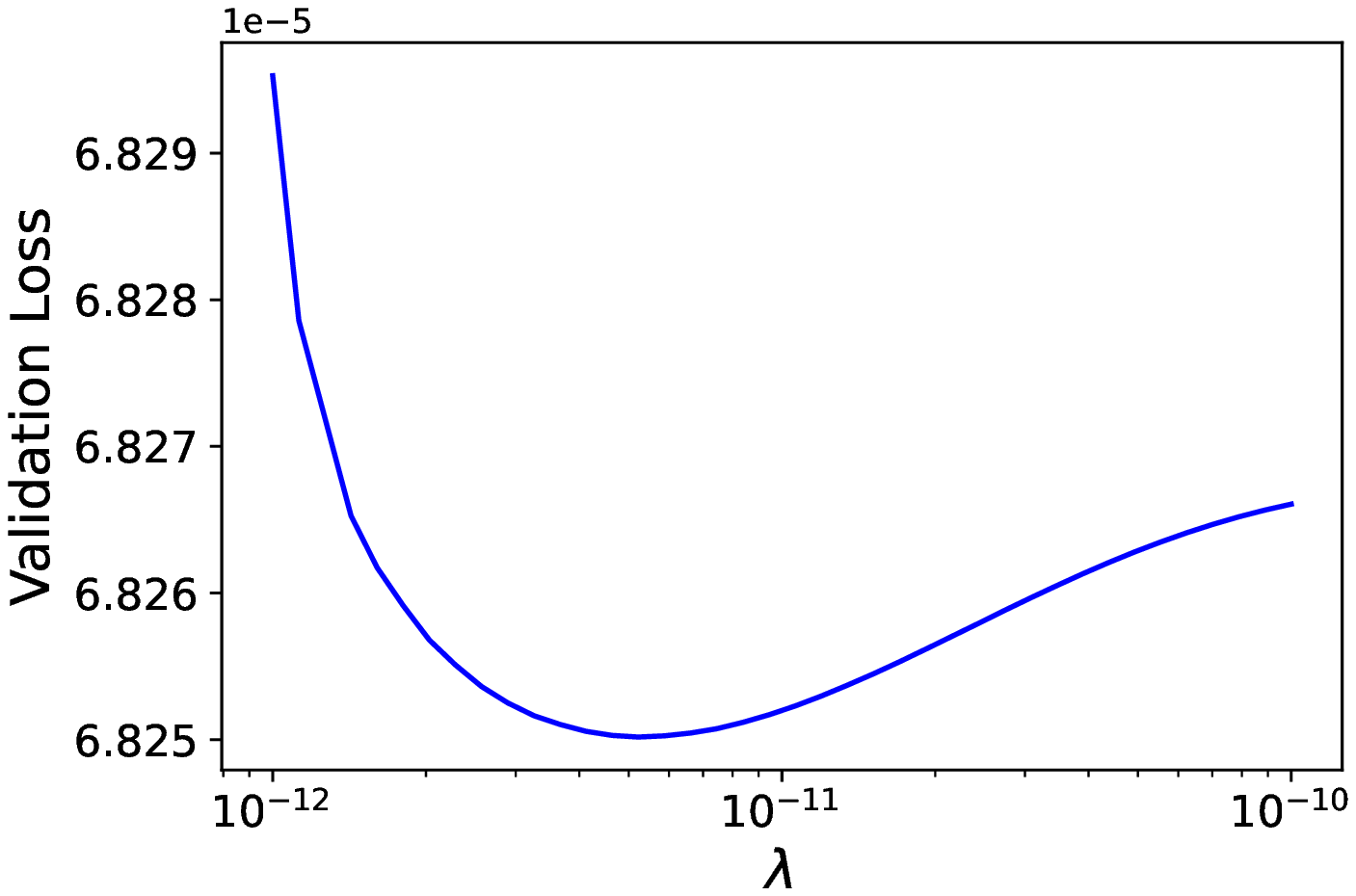}
\caption{Loss computed for the validation set for two systems. Validation loss is plotted against the $\lambda$ value for \ethylene in the STO-3G basis (left) and for \heh in 6-311++G** basis (right). Note the x axis is plotted on a log scale. The $\lambda$ value that minimizes the validation loss is chosen. For \ethylene, $\lambda = 1.1\times 10^{-6}$ and for \heh, $\lambda = 5.2 \times 10^{-12}$. }
\label{fig1}
\end{figure}

When we scale our method to molecular systems with large $N$, we also notice multicollinearity, e.g., numerous zero eigenvalues in the Hessian of the loss $\mathcal{L}$. With multicollinear data, the least squares estimator predicts poorly. We eliminate this problem by using ridge regression, for which we can write the penalized loss function as $\mathcal{L}_{\lambda}( \beta ) =  \mathcal{L}( \beta )  + \lambda \|\beta\|_2^2$; note the use of the $2$-norm, as opposed to the $1$-norm in the penalty term for Lasso, i.e., $\lambda \|\beta\|_1$ \cite{hastie01statisticallearning}. In this work, we train our model by computing the ridge regression solution:  %$\beta_{\text{ridge}} = -(Q +2 \lambda I)^{-1} c^T ,$
\begin{equation}
    \beta_{\text{ridge}} = -\big(Q +2 \lambda I\big)^{-1}c^T,
\end{equation}
where $Q$ is the Hessian of $\mathcal{L}$ with respect to $\beta$ and $c$ is the gradient of $\mathcal{L}$ with respect to $\beta$ computed at $\beta=0$. For a grid of $\lambda$ values, we compute $\beta_{\text{ridge}}$ on the training set, and then compute the loss on a validation set that is disjoint from but equal in size to the training set. Figure \ref{fig1} shows the validation loss for different $\lambda$ values for \ethylene in the STO-3G basis and \heh in the 6-311G** basis set. We choose $\lambda$ that minimizes the validation set loss.

One might conclude incorrectly from Figure \ref{fig1} that, as the range of $\lambda$ values on the vertical axes is small (multiplied by $10^{-2}$ in the left panel and $10^{-5}$ on the right), choosing a non-optimal $\lambda$ may not affect final results. In practice, we find that field-on propagation results improve considerably if we choose the optimal $\lambda$. We hypothesize that this occurs for two reasons.  First, the loss function essentially measures local or one-step propagation error, as described in Section \ref{sect:statlearning}. Second, we note that \eqref{eqn:eq1} is a nonlinear system of ordinary differential equations; the right-hand side is quadratic in the elements of $\mathbf{P}(t)$.  Nonlinearity can magnify errors in the estimated Hamiltonian.  Over thousands of time steps, these errors can accumulate and cause predicted trajectories to diverge substantially from reality.  We also explored Lasso, but chose ridge regression due to its superior performance.

\begin{figure}[t]
\centering
\includegraphics[width=3.2in]{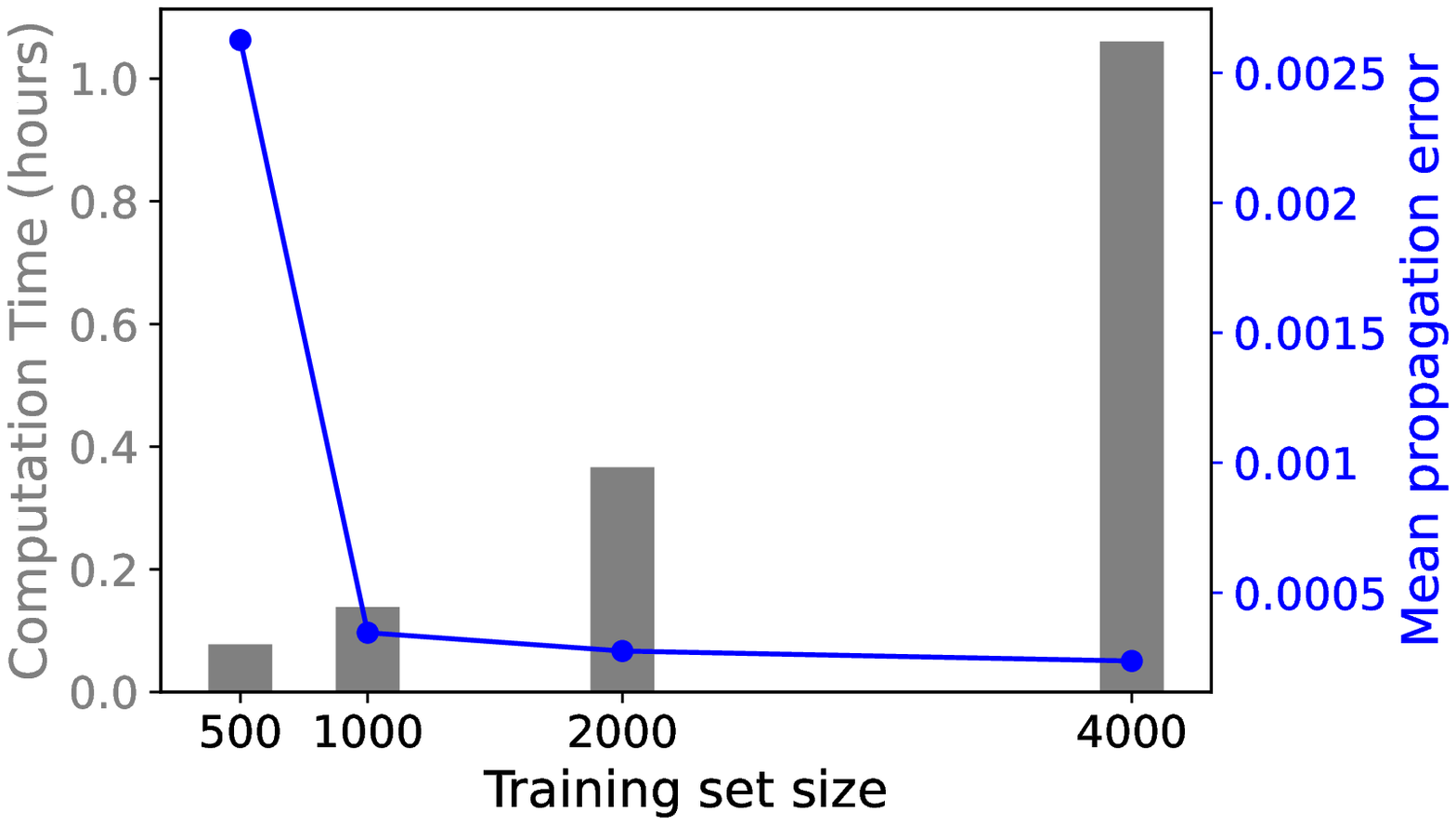}
\includegraphics[width=3.0in]{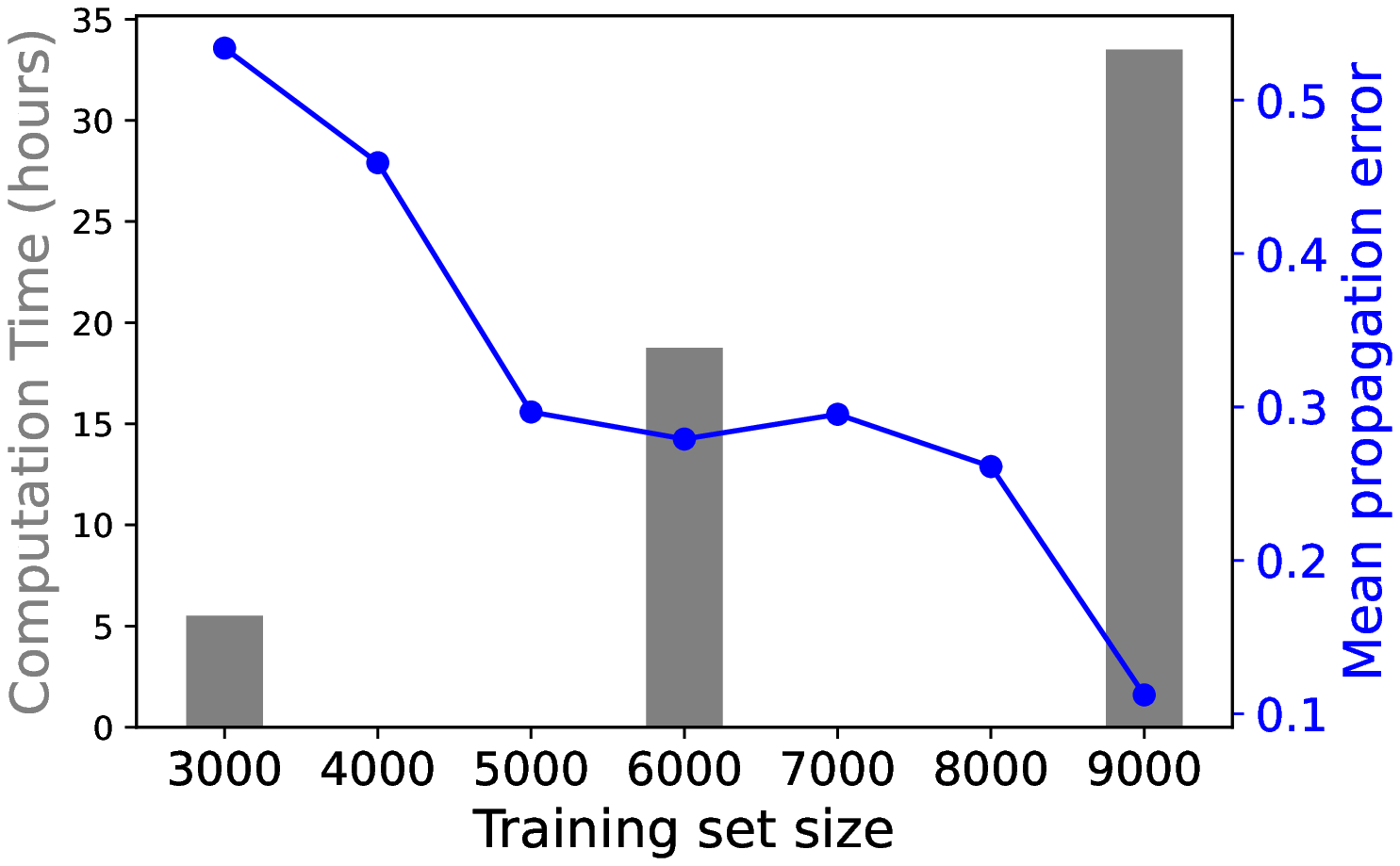}
\caption{Training set size vs. mean propagation error for \heh in 6-311++G$^{**}$ (left) and \lih in 6-311++G$^{**}$ (right). As we increase the training set size the test error decreases but the computational time for training increases.}
\label{fig:learningcurve}
\end{figure}

% \begin{center}
% \begin{table*}[t]%
% \caption{This is sample table caption.\label{tab1}}
% \centering
% \begin{tabular*}{500pt}{@{\extracolsep\fill}lccD{.}{.}{3}c@{\extracolsep\fill}}
% \toprule
% &\multicolumn{2}{@{}c@{}}{\textbf{Spanned heading\tnote{1}}} & \multicolumn{2}{@{}c@{}}{\textbf{Spanned heading\tnote{2}}} \\\cmidrule{2-3}\cmidrule{4-5}
% \textbf{col1 head} & \textbf{col2 head}  & \textbf{col3 head}  & \multicolumn{1}{@{}l@{}}{\textbf{col4 head}}  & \textbf{col5 head}   \\
% \midrule
% col1 text & col2 text  & col3 text  & 12.34  & col5 text\tnote{1}   \\
% col1 text & col2 text  & col3 text  & 1.62  & col5 text\tnote{2}   \\
% col1 text & col2 text  & col3 text  & 51.809  & col5 text   \\
% \bottomrule
% \end{tabular*}
% \begin{tablenotes}%%[341pt]
% \item Source: Example for table source text.
% \item[1] Example for a first table footnote.
% \item[2] Example for a second table footnote.
% \end{tablenotes}
% \end{table*}
% \end{center}

% Below is the example for bulleted list. Below is the example for bulleted list. Below is the example for bulleted list\footnote{This is an example for footnote.}:
% \begin{itemize}

% \item sample list entry text. sample list entry text.  
% \end{itemize}

\section{Results}\label{sec4}

% , showing that we can learn the Hamiltonian and then simulate the electron dynamics for arbitrary field perturbations.  This increase in the system size leads to multicollinearity and sparsity, and dimensionality reduction based on non-zero elements of the density matrix is not sufficient.   
Applying the training procedure described in Section \ref{sec2} to the molecular systems listed in Table \ref{tab1}, we learn $\beta$ and determine $\widetilde{\mathbf{H}}$. Here, for smaller molecular systems, we train using time series with $2000$ points. For larger systems, we increase the training set size; we determine the number of points by computing a learning curve, plotting test set propagation error against the number of training points. Let us illustrate the effect of training set size for two of the larger systems studied here: \heh in 6-311++G** and the largest molecular system, \lih in 6-311++G$^{**}$. Figure \ref{fig:learningcurve} shows that, as we increase the training set size, field-on propagation error decreases (blue) while computational time for training increases (gray).  The training set size used for each system is given in \ref{tab1}.

For propagation, we use RK45 (\cite{dormand1980}) to solve \eqref{eqn:eq1} numerically with the learned Hamiltonian $\widetilde{\mathbf{H}}$ for 2000 steps.  We do this both for the case of a delta kick perturbation (the same as the training data, a \emph{field-off} perturbation) and for the case of a sinusoidal electric field perturbation (a \emph{field-on} perturbation). The field-on perturbation tests the learned Hamiltonian in a regime that is outside that of the training set. 

\begin{table}[h]%
\centering
\small
\caption{Molecule, number of elements in the density matrix, training loss, field-free and field-on propagation error.\label{tab1}}%
\begin{tabular}{lccccccc}
\hline
\textbf{Molecule} &\textbf{Basis set}  &\textbf{$\mathbf{N^2}$}&\textbf{Training Set size} &${\mathbf{\lambda}}$  & \textbf{Training Loss}  & \textbf{field-free error}  & \textbf{field-on error} \\
\hline
\htwo & 6-31G &$16$&$1000$&$0$ & $7.15 \times 10^{-6}$& $3.09 \times 10^{-3}$ & $6.31 \times 10^{-4}$ \\
\heh & 6-31G &$16$&$2000$&$0$ &$8.99 \times 10^{-5}$ & $6.50 \times 10^{-3}$ & $2.53 \times 10^{-4}$\\
\lih &6-31G &$121$&$2000$& $1.0 \times 10^{-8}$ &$1.39 \times 10^{-5}$ & $6.82\times 10^{-3}$& $6.01\times 10^{-3}$\\
\ethylene & STO-3G& $196$& $2000$& $1.1 \times 10^{-6}$   &$2.72 \times 10^{-2}$  &$5.22 \times 10^{-2}$& $1.38 \times 10^{-3}$ \\
\heh & 6-311++G**&$196$& $4000$&$5.2 \times 10^{-12}$  & $4.68 \times 10^{-5}$ &$8.84 \times 10^{-3}$ & $3.02 \times 10^{-4}$\\
\lih &6-311++G** &$841$&$9000$ & $5.0 \times 10^{-6}$ & $4.79 \times 10^{-5}$& $1.52 \times 10^{-2}$& $1.71 \times 10^{-1}$\\
\hline
\end{tabular}

\end{table}

The training loss, field-free, and field-on propagation error for six molecular systems are presented in Table \ref{tab1}. Training loss reported here is calculated as $\mathcal{L}(\beta^*)$ using  \eqref{eqn:sseloss}. This training loss measures the squared Frobenius norm of one step errors, i.e, the error in propagating to the next time step using the learned Hamiltonian via (\ref{eqn:onesteppred}). The small values of the field-free error, for all molecules, indicate that the Hamiltonian learned by minimizing \eqref{eqn:sseloss} can be used for long-term propagation. Even with an applied field, which is outside the training regime, we obtain propagation errors comparable to if not less than those in the field-free case, implying that the learned Hamiltonian generalizes well beyond the training regime.
\begin{figure}[h!]
\centering
\includegraphics[width=5.5in]{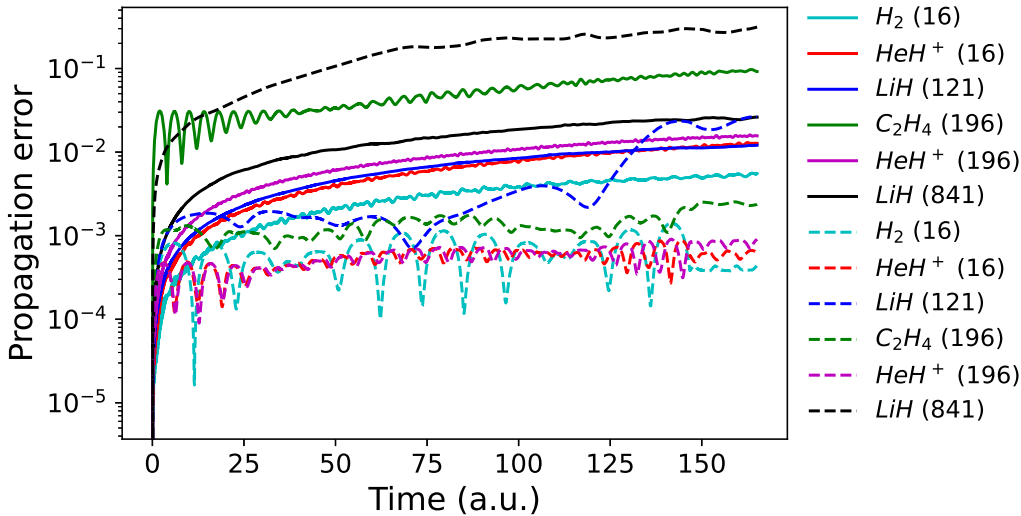}
\caption{Propagation error compares ground truth density matrices against those computed by numerically solving (\ref{eqn:eq1}) using the learned Hamiltonian $\widetilde{\mathbf{H}}$. The solid lines are for field-off propagation and the dashed lines are with the field on.\label{fig:Errorprop}}
\end{figure}

Let $\mathbf{P^\prime}$ denote the prediction, i.e, density matrix obtained by propagating the learned Hamiltonian. We define the time-dependent propagation error as
\begin{equation}
\mathcal{E }(t_j) = \|\mathbf{P}^\prime (t_j)-\mathbf{P}(t_j) \|_F, 
    \label{eqn:TDproperror}
\end{equation}
where $\mathcal{E }(t_j)$ measures the error (at time $t_j$) between $\mathbf{P}^\prime$, the predicted trajectory obtained by propagating the learned Hamiltonian, and $\mathbf{P}$, the ground truth trajectory. 
We calculate the mean propagation error for the propagation interval as 
\begin{equation}
\mathcal{E} = \frac{1}{M}\sum_{j=1}^{M}\mathcal{E }(t_j),  
    \label{eqn:properror}
\end{equation}
where $M$ is the number of time steps for which we propagate the Hamiltonian. For this study, $M=2000$. In Fig. \ref{fig:Errorprop} we plot the time-dependent propagation errors $\mathcal{E}(t_j)$ for all molecular systems in both the field-free and field-on cases. We see that that errors for both cases remain reasonably small for all molecular systems even after propagating for $150$ a.u., which is equivalent to $2000$ time steps. 

\begin{figure}[htbp]
\begin{center}
\includegraphics[width=3.0in]{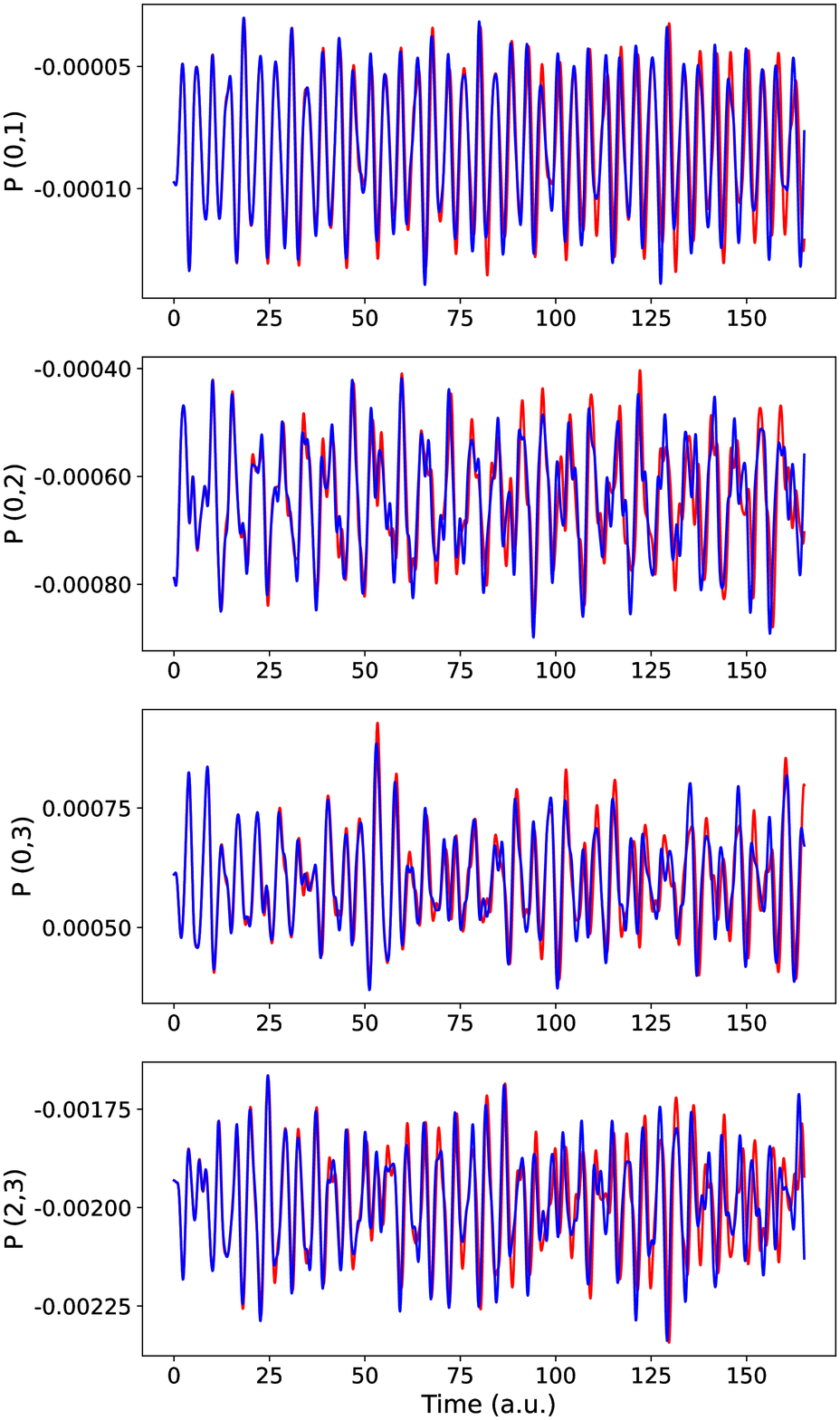}
\includegraphics[width=3.0in]{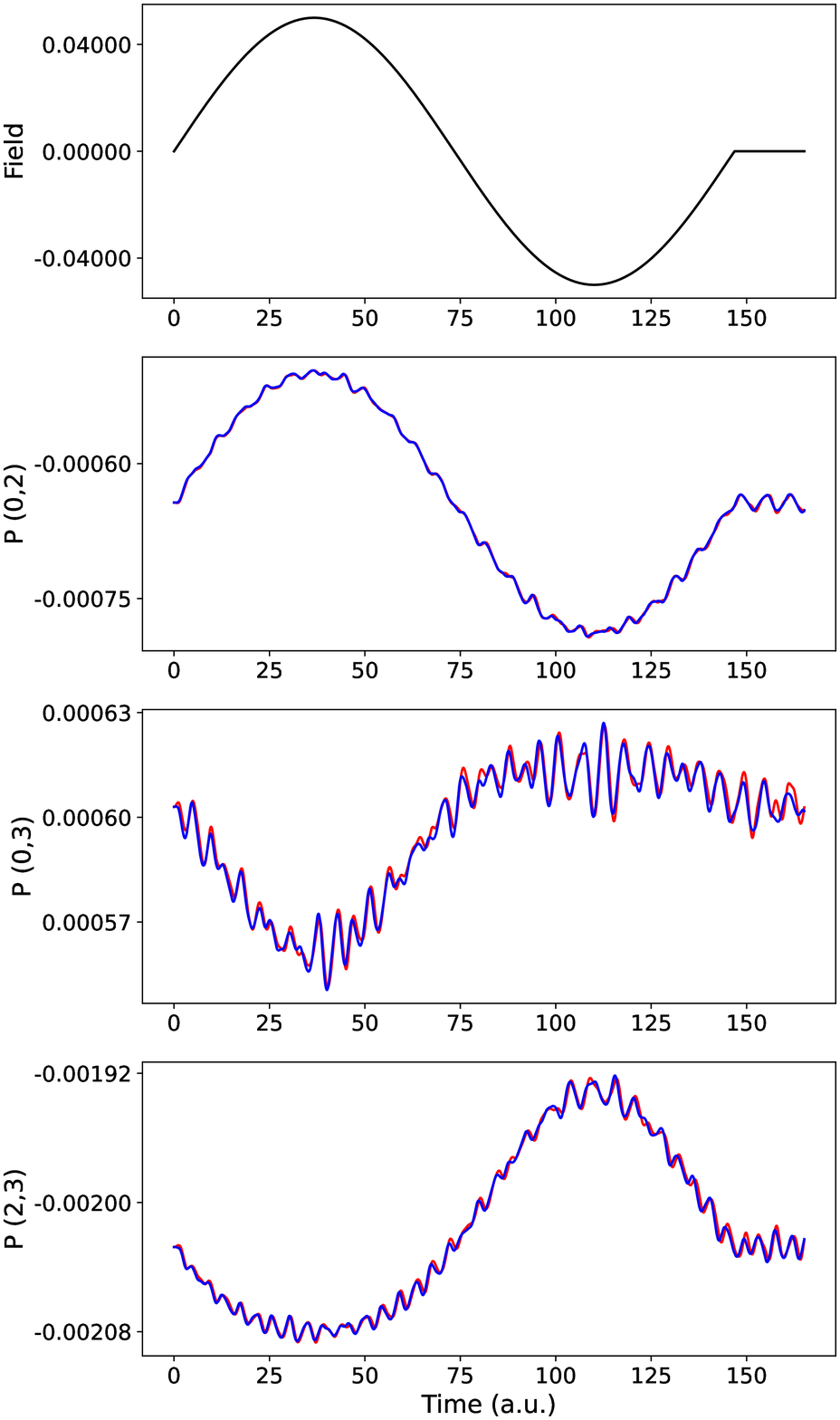}
\end{center}
\caption{Real parts of selected elements of ground truth density matrices (blue) and density matrices computed using the learned Hamiltonian $\widetilde{\mathbf{H}}$ (red) for \heh in the 6-311++G** basis for the field-free (left) and field-on (right) cases.  Note the close agreement between all curves.\label{fig4}}
\end{figure}

In Fig. \ref{fig4}, we plot, as a function of time, selected nonzero elements of the density matrix obtained by propagating the learned Hamiltonian (red), and the ground truth (blue) obtained from a widely-used electronic structure code (see details in Section \ref{sec2}). We observe good agreement between predicted and ground truth trajectories.

% \begin{figure*}
% \centering
% \begin{minipage}[b]{0.4\textwidth}
%     \includegraphics[width=\columnwidth]{heh+prop}
%     %\caption{a.}
%   \end{minipage}
% \centering
% \begin{minipage}[b]{0.4\textwidth}
%     \includegraphics[width=\columnwidth]{hehpropWF}
%     %\caption{b.}
%   \end{minipage}

% \caption{Selected elements of ground truth density matrices (black), density matrices computed using the learned Hamiltonian $\widetilde{\mathbf{H}}$ (red), and density matrices computed using the exact  Hamiltonian $\mathbf{H}$ (blue) for \heh in the 6-311++G** basis for field free and field on.  Note the close agreement between all curves.\label{fig4}}
% \end{figure*}

% \begin{equation}\label{eq23}
%  \|\tilde{X}(k)\|^2
%  =\frac{\left\|\sum\limits_{i=1}^{p}\tilde{Y}_i(k)+\sum\limits_{j=1}^{q}\tilde{Z}_j(k) \right\|^2}{(p+q)^2}
%  \leq\frac{\sum\limits_{i=1}^{p}\left\|\tilde{Y}_i(k)\right\|^2+\sum\limits_{j=1}^{q}\left\|\tilde{Z}_j(k)\right\|^2 }{p+q}.
% \end{equation}

% \begin{equation}\label{eq24}
%  \|\tilde{X}(k)\|^2
%  =\frac{\left\|\sum\limits_{i=1}^{p}\tilde{Y}_i(k)+\sum\limits_{j=1}^{q}\tilde{Z}_j(k) \right\|^2}{(p+q)^2}
%  \leq\frac{\sum\limits_{i=1}^{p}\left\|\tilde{Y}_i(k)\right\|^2+\sum\limits_{j=1}^{q}\left\|\tilde{Z}_j(k)\right\|^2 }{p+q}.
% \end{equation}

\section{Discusssion}
\label{sec5}
In this work, we extended our prior methodology by incorporating dimensionality reduction (in the form of real-imaginary splitting) and ridge regression.  Using these techniques, we addressed challenges in scaling our method to molecular systems with larger basis set size $N$. Using the learned Hamiltonian, we can predict electron densities for not only the training set (field free) but also for the test set (field on). The loss function \eqref{eqn:sseloss} measures the sum of squares of one-step propagation errors, a form of local error. By minimizing this loss over the training set, we obtain very good long-time propagation error. For some molecular systems, the agreement is to a degree that we cannot tell the two curves (propagation using learned Hamiltonian and the ground truth trajectory) apart.

We used two dimensionality reduction techniques to significantly reduce the number of model parameters: (i) splitting the Hamiltonian model based on properties of the HF Hamiltonian and (ii) modeling only non-zero elements of the Hamiltonian matrix. The effective number of degrees of freedom in the Hamiltonian is less than the number of non-zero elements due to the linear combinations of Hamiltonian elements that  expressed through \eqref{eqn:eq1}. This reduction can be easily observed for smaller molecular systems like \htwo in the STO-3G basis set. For larger molecular systems, these linear dependencies are much more prevalent and more difficult to verify directly. In such cases, regularization improves the prediction capability of a model by decreasing the number of degrees of freedom. Here, using ridge regression we successfully reduced the field-on propagation error.

We also coded the Hessian and gradient for the loss function instead of using automatic differentiation techniques, thus making it feasible to obtain the least squares solution for larger molecular systems. Although for most molecular systems we used one field-free trajectory with $2000$ time steps for training, for larger systems such as \lih ($N^2 = 841$), we increased the training set size and observed that the field-on propagation error decreases for a larger training set. However, as we increase the training set size, the computational time for training increases, eventually becoming prohibitively expensive for a training set with more than $9000$ time steps.  In the future, we hope to extend this model to even larger molecular systems, and also learn a density-dependent Hamiltonian based on more accurate wave function generated densities.
%\backmatter

\section*{Acknowledgments}
This work was supported by the U.S. Department of Energy, Office of Science, Office of Basic Energy Sciences under Award Number DE-SC0020203.  We acknowledge computational time on the MERCED cluster (funded by NSF ACI-1429783), and on the Nautilus cluster, which is supported by the Pacific Research Platform (NSF ACI-1541349), CHASE-CI (NSF CNS-1730158), and Towards a National Research Platform (NSF OAC-1826967). Additional funding for Nautilus has been supplied by the University of California Office of the President.

\subsection*{Data Availability Statement}

All code required to reproduce all training and test results is available on GitHub at https:
\url{//github.com/hbhat4000/electrondynamics} \cite{BhatGitRepo}. Training data is available from the authors upon request.

\subsection*{Financial disclosure}

None reported.

\subsection*{Conflict of interest}

The authors declare no potential conflict of interests.

%\section*{Supporting information}

%The following supporting information is available as part of the online article:

% \noindent
% \textbf{Figure S1.}
% {500{\uns}hPa geopotential anomalies for GC2C calculated against the ERA Interim reanalysis. The period is 1989--2008.}

% \noindent
% \textbf{Figure S2.}
% {The SST anomalies for GC2C calculated against the observations (OIsst).}

% \begin{align*}
%     \mathcal{L}(\beta) &= \|y-X\beta\|^2_2 \\
%     &= (y-X\beta )^{T}(y-X\beta) \\
%     &= y^Ty -y^TX\beta - \beta^TX^Ty+\beta^TX^TX\beta
% \end{align*}

\appendix

\section{Reduction to Least Squares}
\label{app1}
We begin by writing the loss (\ref{eqn:sseloss}) as $\mathcal{L}(\beta) = \|y-X\beta\|^2_2$. To minimize $\mathcal{L}$, we need to determine
\begin{equation}
    \beta^* \in \underset{\beta}{\arg\min} \{\mathcal{L}(\beta)\}.
\end{equation}
We start by expanding the loss function: 
\begin{equation*}
    \mathcal{L}(\beta) = \|y-X\beta\|^2_2 = y^Ty -y^TX\beta - \beta^TX^Ty+\beta^TX^TX\beta.
\end{equation*}
To minimize the right-hand side, we take the gradient with respect to $\beta$,
\begin{equation*}
\nabla_{\beta} L(\beta) = -2X^T y + 2 X^T X \beta.
\end{equation*}
Setting this gradient to $0$, we obtain the normal equations:
\begin{equation}
    2 X^T X \beta^\ast = 2X^T y.
    \label{eqA2}
\end{equation}
Let $H_{\beta} \mathcal{L}$ denote the Hessian of $\mathcal{L}$.  Since $\nabla_{\beta} \mathcal{L}(0) = -2X^T y$ and  $H_{\beta}\mathcal{L}=  2X^T X $, we can write (\ref{eqA2}) as
\begin{equation}
    (H_{\beta}\mathcal{L}) \beta^\ast = -\nabla_{\beta} \mathcal{L}(0)
    \label{eqA3}
\end{equation}
To estimate the ridge regression solution, we need to compute
\begin{equation}
    \beta_{\text{ridge}}^* \in \underset{\beta}{\arg\min} \{\mathcal{L}(\beta)\ + \lambda \|\beta\|^2_2\}.
\end{equation}
Augmenting the loss with the ridge penalty yields
\begin{equation*}
 \mathcal{L}_{\lambda}(\beta) =  \|y-X\beta\|^2_2 + \lambda \| \beta\|_2^2 = y^Ty -2 y^TX\beta+\beta^TX^TX\beta + \lambda \beta^T \beta
\end{equation*}
The gradient is then
\begin{equation*}
\nabla_{\beta} L_{\lambda}(\beta) = -2 X^T y+ 2 X^T X \beta+2 \lambda \beta
\end{equation*}
Setting the gradient to $0$, we get
\begin{equation}
    (2 X^T X + 2\lambda I) \beta_{\text{ridge}}^\ast = 2X^T y.
    \label{eqA5}
\end{equation}
The Hessian of $\mathcal{L}^\lambda$ is $H_{\beta} \mathcal{L}^\lambda =  2X^T X + 2\lambda I $.  With this, we can write (\ref{eqA5}) as
\begin{equation}
    (H_{\beta}\mathcal{L} + 2\lambda I) \beta_{\text{ridge}}^\ast = - \nabla_{\beta} \mathcal{L}(0) 
\end{equation}

\section{Computation of the Gradient and Hessian}
\label{sect:gradHess}
Here we describe the details behind our computation of the gradient and Hessian of the loss function \eqref{eqn:sseloss}. 
Let us introduce the notation $P^j_{mn}$ to denote the $m$-th row and $n$-th column of the matrix $\mathbf{P} = \mathbf{P}(t_j)$. Similarly, let $\dot{P}^j_{mn}$ denote the $m$-th row and $n$-th column of the centered-difference time derivative $\dot{\mathbf{P}} = (\mathbf{P}(t_{j+1}) - \mathbf{P}(t_{j-1}))/(2 \Delta t)$.  We let $\mathbf{H}$ denote $\widetilde{\mathbf{H}}(\mathbf{P}(t_j))$.  Then, with $\ast$ denoting complex conjugate in this section, we can rewrite the loss \eqref{eqn:sseloss} as
\begin{equation}
\label{eqn:newloss}
\mathcal{L}( \beta ) =  \sum_{m,n} \mathcal{L}_{mn}(\beta), \text{ where } \mathcal{L}_{mn}(\beta) = \sum_{j=1}^{N-1} \biggl| i  \dot{P}^j_{mn}
    -  \big [  \mathbf{H} , \mathbf{P} \big]^j_{mn} \biggr|^2 = \sum_{j=1}^{N-1} \biggl( i  \dot{P}^j_{mn}
    -  \big [  \mathbf{H} , \mathbf{P} \big]^j_{mn} \biggr) \biggl( -i  \dot{P}^{j \ast}_{mn}
    -  \big [  \mathbf{H} , \mathbf{P} \big]^{j \ast}_{mn} \biggr).
\end{equation}
Whereas we previously wrote $\beta = (\beta_0, \beta_1)$, here we give more details.  The term $\beta_0$ refers to an intercept matrix.  However, $\beta_1$ refers to two collections of matrices, $\{\eta_k\}_{k=1}^K$ and $\{\gamma_{\ell}\}_{\ell=1}^L$.  All matrices here are of the same dimension as $\mathbf{H}$.  To better understand the roles of these matrices, let us note that $\mathbf{H}$ depends only on certain non-zero, upper-triangular entries of $\mathbf{P}$.  We let $\{ r_1, \ldots, r_K\}$ denote the \emph{indices} of the real part of $\mathbf{P}$ upon which we allow $\mathbf{H}$ to depend.  Similarly, we let $\{i_1 , \ldots, i_L\}$ denote the \emph{indices} of the imaginary part of $\mathbf{P}$ upon which we allow $\mathbf{H}$ to depend.  Hence $K$ and $L$ are, respectively, the total numbers of real and imaginary parts of $\mathbf{P}$ that are \emph{active} in the model for $\mathbf{H}$.

With this notation, we can write our linear model for $\mathbf{H}$ as follows---note that we begin with the upper-triangular part: for $m \leq q$,
\begin{equation}
\label{eqn:Hupper}
H_{mq} = \beta_0^{mq} + \sum_{k=1}^K P^j_{r_k} \eta_k^{mq} + i \sum_{\ell=1}^L P^j_{i_{\ell}} \gamma_{\ell}^{mq}.
\end{equation}
For $m > q$, because $\mathbf{H}$ is Hermitian (or self-adjoint), we have
\begin{equation}
\label{eqn:Hlower}
H_{mq} = H_{qm}^{\ast} = \beta_0^{qm\ast} + \sum_{k=1}^K P^{j\ast}_{r_k} \eta_k^{qm} - i \sum_{\ell=1}^L P^{j\ast}_{i_{\ell}} \gamma_{\ell}^{qm}.
\end{equation}
Here we have used the fact that $\eta$ and $\gamma$ are both real---this is necessary for the real (respectively, imaginary) part of $\mathbf{H}$ to depend only on the real (respectively, imaginary) part of $\mathbf{P}$.  Note that in these expressions, we only use the upper-triangular parts of $\eta$ and $\gamma$.

We focus first on the gradient and Hessian of $\mathcal{L}_{mn}$ with respect to $\eta_{k}$.  From (\ref{eqn:Hupper}-\ref{eqn:Hlower}), we see that $\mathcal{L}_{mn}$ depends on $\beta$ only through $\mathbf{H}$.  For any integers $j$ and $k$, we define the Kronecker delta $\delta_{jk} = \begin{cases} 1 & j = k \\ 0 & j \neq k \end{cases}$.  Then
\[
\frac{\partial H_{mq}}{\partial \eta_s^{tu}}  = \begin{cases} P_{r_s}^j \delta_{tm} \delta_{uq} & m \leq q \\
 P_{r_s}^{j\ast} \delta_{tq} \delta_{um} & m > q. \end{cases}
\]
Observe that $\mathcal{L}_{mn}$ is of the form $\sum_{j} Z_{mn} Z_{mn}^\ast$.  Putting these pieces together, we obtain, with $\Re$ signifying real part,
\begin{equation}
\label{eqn:lossmnpart}
\frac{\partial \mathcal{L}_{mn}}{\partial \eta_{s}^{tu}} = 2 \Re \sum_{j} \left( \frac{\partial}{\partial \eta_{s}^{tu}} \big [  \mathbf{H} , \mathbf{P} \big]^j_{mn} \right) \biggl( i  \dot{P}^{j \ast}_{mn}
    + \big [  \mathbf{H} , \mathbf{P} \big]^{j \ast}_{mn} \biggr).
\end{equation}
In what follows, we use $I_A$ to denote the indicator function of the set $A$, {\it e.g.}, $I_{j>k} = \begin{cases} 1 & j > k \\ 0 & j \leq k\end{cases}$. We then compute
\begin{align*}
\frac{\partial}{\partial \eta_{s}^{tu}} \big [  \mathbf{H} , \mathbf{P} \big]^j_{mn} &= \frac{\partial}{\partial \eta_{s}^{tu}} \left( \sum_{q} H_{mq} P^j_{qn} - P^j_{mq} H_{qn} \right) \\
 &= \sum_q \left( P_{r_s}^j \delta_{tm} \delta_{uq} I_{q \geq m} + P_{r_s}^{j\ast} \delta_{tq} \delta_{um} I_{q < m} \right) P^j_{qn} - P^j_{mq} \left( P_{r_s}^j \delta_{tq} \delta_{un} I_{q \leq n} + P_{r_s}^{j\ast} \delta_{tn} \delta_{uq} I_{q > n} \right) \\
 &= P_{r_s}^j \delta_{tm} P^j_{un} I_{u \geq m} + P_{r_s}^{j\ast} \delta_{um} P^j_{tn}  I_{t < m} - P^j_{mt} P_{r_s}^j  \delta_{un} I_{t \leq n} - P_{r_s}^{j\ast} P_{mu}^j \delta_{tn} I_{u>n}
\end{align*}
Hence
\[
\frac{\partial \mathcal{L}_{mn}}{\partial \eta_{s}^{tu}} = 2 \Re \sum_{j} \left( P_{r_s}^j \delta_{tm} P^j_{un} I_{u \geq m} + P_{r_s}^{j\ast} \delta_{um} P^j_{tn}  I_{t < m} - P^j_{mt} P_{r_s}^j  \delta_{un} I_{t \leq n} - P_{r_s}^{j\ast} P_{mu}^j \delta_{tn} I_{u>n} \right) \biggl( i  \dot{P}^{j \ast}_{mn}
    + \big [  \mathbf{H} , \mathbf{P} \big]^{j \ast}_{mn} \biggr).
\]
This implies that
\begin{subequations}
\label{eqn:gradeta}
\begin{align}
\frac{\partial \mathcal{L}}{\partial \eta_{s}^{tu}} &= \sum_{m,n} \frac{\partial \mathcal{L}_{mn}}{\partial \eta_{s}^{tu}} \nonumber \\
&= 2 \Re \sum_{m,n} \sum_{j} \left( P_{r_s}^j \delta_{tm} P^j_{un} I_{u \geq m} + P_{r_s}^{j\ast} \delta_{um} P^j_{tn}  I_{t < m} - P^j_{mt} P_{r_s}^j  \delta_{un} I_{t \leq n} - P_{r_s}^{j\ast} P_{mu}^j \delta_{tn} I_{u>n} \right) \biggl( i  \dot{P}^{j \ast}_{mn}
    + \big [  \mathbf{H} , \mathbf{P} \big]^{j \ast}_{mn} \biggr) \nonumber \\
    &= 2 \Re \biggl[ \sum_{j,n} \left\{ P_{r_s}^j P_{un}^j \left( i \dot{P}^{j\ast}_{tn} + [H,P]^{j\ast}_{tn} \right) I_{u \geq t}
     + P_{r_s}^{j \ast} P_{tn}^j  \left( i \dot{P}^{j\ast}_{un} + [H,P]^{j\ast}_{un} \right) I_{u > t} \right\} \\
  &\quad \quad - \sum_{j,m} \left\{ P_{r_s}^j P_{mt}^j \left( i \dot{P}^{j\ast}_{mu} + [H,P]^{j\ast}_{mu} \right) I_{u \geq t}
    + P_{r_s}^{j \ast} P_{mu}^j \left( i \dot{P}^{j \ast}_{mt} + 
     [H,P]^{j \ast}_{mt} \right) I_{u>t} \right\} \biggr].
\end{align}
\end{subequations}
This is the gradient of the loss with respect to each of the $\eta_s$ matrices.  In our code, we parallelize this computation across the $t$ and $u$ indices.  More specifically, we implement this calculation via a function that, for a given $t$ and $u$, computes $\partial \mathcal{L} / \partial \eta_s^{tu}$ for all $s$ at once.  We then evaluate this function in parallel across all indices $t \leq u$; as mentioned above, the lower-triangular parts of the $\beta_0$, $\eta$, and $\gamma$ matrices play no role in our model for $\mathbf{H}$.

Examining the form of the model (\ref{eqn:Hupper}-\ref{eqn:Hlower}), we note that upon exchanging
\begin{equation}
\label{eqn:swap}
P_{r_s}^j \longleftrightarrow i P_{i_s}^j \quad \text{ and } \quad P_{r_s}^{j\ast} \longleftrightarrow -i P_{i_s}^{j\ast},
\end{equation}
the roles of $\eta$ and $\gamma$ become reversed.  Using this fact, we can extract from the above calculation an expression for the gradient $\partial \mathcal{L} / \partial \gamma_s^{tu}$: we simply apply the transformation (\ref{eqn:swap}) to (\ref{eqn:gradeta}).  We have verified by hand that this yields precisely the same result as differentiating $\mathcal{L}$ directly with respect to $\gamma_s^{tu}$.

Further examining (\ref{eqn:Hupper}-\ref{eqn:Hlower}), we see that if we set $K=1$ and $P^j_{r_k} \equiv 1$, then $\eta$ plays the same role as $\beta_0$.  Setting $P^j_{r_k} \to 1$ in (\ref{eqn:gradeta}) gives us  the gradient $\partial \mathcal{L} / \partial \beta_0^{tu}$; again, we have verified this by hand.  With this, we have described the full computation of the gradient of $\mathcal{L}$ with respect to all model parameters.

To begin our calculation of the Hessian, we take a second $\eta$ derivative on both sides of (\ref{eqn:lossmnpart}) to obtain
\begin{align*}
\frac{\partial \mathcal{L}_{mn}}{\partial \eta_{s}^{tu} \partial \eta_{a}^{bc}} &= 2 \Re \sum_{j} \left( \frac{\partial}{\partial \eta_{s}^{tu}} \big [  \mathbf{H} , \mathbf{P} \big]^j_{mn} \right) \left( \frac{\partial}{\partial \eta_{s}^{tu}} \big [  \mathbf{H} , \mathbf{P} \big]^{j\ast}_{mn} \right) \\
 &= 2 \Re \sum_j (P_{r_s}^j \delta_{tm} P^j_{un} I_{u \geq m} + P_{r_s}^{j\ast} \delta_{um} P^j_{tn}  I_{t < m} - P^j_{mt} P_{r_s}^j  \delta_{un} I_{t \leq n} - P_{r_s}^{j\ast} P_{mu}^j \delta_{tn} I_{u>n})\\
 &\quad \quad \quad \quad \ \cdot 
 (P_{r_a}^{j\ast} \delta_{bm} P^{j\ast}_{cn} I_{c \geq m} + P_{r_a}^{j} \delta_{cm} P^{j\ast}_{bn}  I_{b < m} - P^{j\ast}_{mb} P_{r_a}^{j\ast}  \delta_{cn} I_{b \leq n} - P_{r_a}^{j} P_{mc}^{j\ast} \delta_{bn} I_{c>n}).
\end{align*}
The product here yields $16$ different terms inside the sum.  Through algebra analogous to that used to derive (\ref{eqn:gradeta}), we can compute each of these $16$ terms and sum them over all $m$ and $n$.  The resulting $16$ terms give us a closed-form expression for $\partial \mathcal{L}/(\partial \eta_{s}^{tu} \partial \eta_{a}^{bc})$.  When we implement the Hessian in code, we first develop a function that takes as input fixed values of $t$, $u$, $b$, and $c$, returning as output the partial derivative $\partial \mathcal{L}/(\partial \eta_{s}^{tu} \partial \eta_{a}^{bc})$ for all values of $s$ and $a$ at once.  We then evaluate this function in parallel over all possible values of $t \leq u$ and $b \leq c$, again taking into account the fact that only the upper-triangular part of $\eta$ matters.  In this way, we compute the central $(2,2)$ block in the overall Hessian:
\begin{equation}
\label{eqn:overallhess}
H_{\beta} \mathcal{L} = \begin{bmatrix} \partial_{\beta_0} \partial_{\beta_0} \mathcal{L} & \partial_{\beta_0} \partial_{\eta} \mathcal{L} & \partial_{\beta_0} \partial_{\gamma} \mathcal{L} \\
 \partial_{\eta} \partial_{\beta_0} \mathcal{L} & \partial_{\eta} \partial_{\eta} \mathcal{L} & \partial_{\eta} \partial_{\gamma} \mathcal{L} \\
 \partial_{\gamma} \partial_{\beta_0} \mathcal{L} & \partial_{\gamma} \partial_{\eta} \mathcal{L} & \partial_{\gamma} \partial_{\gamma} \mathcal{L} \end{bmatrix}.
\end{equation}
The calculation of the $(2,2)$ block can be recycled and converted into calculations of all other blocks.  For instance, applying the transformation (\ref{eqn:swap}) to $P_{r_s}^j$ in the final expression of the $(2,2)$ block gives us, by symmetry, both the $(2,3)$ and $(3,2)$ blocks.  If we then go back and apply the transformation (\ref{eqn:swap})to both $P_{r_s}^j$ \emph{and} $P_{r_a}^j$ in the final expression of the $(2,2)$ block, we obtain the $(3,3)$ block.  Similarly, setting either or both of $\left\{ P_{r_s}^j, P_{r_a}^j \right\}$ to $1$ yields the blocks in the first row and first column of (\ref{eqn:overallhess}).

Through these strategies, we compute all entries of the gradient and Hessian of $\mathcal{L}$ without recourse to automatic differentiation, which we relied upon in our earlier work \cite{bhat2020machine}.  While automatic differentiation yields perfectly accurate gradients and Hessians for small molecular systems, as the system size grows larger, we find that the computational cost of automatic differentiation increases considerably until it becomes unusable.  Simultaneously, we find that the  parallel computation of analytically derived gradients and Hessians, via the techniques described here, scales well to all molecular systems described in this paper.

%\nocite{*}% Show all bib entries - both cited and uncited; comment this line to view only cited bib entries;
\bibliographystyle{abbrvnat}
\bibliography{wileyNJD-APA}%

% \section*{Author Biography}

% \begin{biography}{\includegraphics[width=60pt,height=70pt,draft]{empty}}{\textbf{Author Name.} This is sample author biography text this is sample author biography text this is sample author biography text this is sample author biography text this is sample author biography text this is sample author biography text this is sample author biography text this is sample author biography text this is sample author biography text this is sample author biography text this is sample author biography text this is sample author biography text this is sample author biography text this is sample author biography text this is sample author biography text this is sample author biography text this is sample author biography text this is sample author biography text this is sample author biography text this is sample author biography text this is sample author biography text.}
% \end{biography}

\end{document}